\documentclass[modern]{aastex61}
\usepackage{verbatim}

\newcommand{\thisevent}{OGLE-2007-BLG-224}

\begin{document}

\shorttitle{OB07224 Confirming Terrestrial Parallax}
\shortauthors{Shan, Y. et al.}


\title{OGLE-2007-BLG-224L: Confirmation of Terrestrial Parallax}

\author{Yutong Shan}
\affiliation{Institut f\"ur Astrophysik, Georg-August-Universit\"at, Friedrich-Hund-Platz 1, 37077 G\"ottingen, Germany}
\author{Jennifer C. Yee}
\affiliation{Center for Astrophysics $|$ Harvard \& Smithsonian, 60 Garden St., Cambridge, MA 02138}
\author{Vanessa P. Bailey}
\affiliation{Jet Propulsion Laboratory, 4800 Oak Grove Dr, M/S 321-123, Pasadena, CA 91109}
\author{Laird M. Close}
\affiliation{Steward Observatory, University of Arizona, 933 North Cherry Avenue, Tucson, AZ 85721, USA}
\author{Phil M. Hinz}
\affiliation{Department of Astronomy and Astrophysics, University of California, Santa Cruz, 1156 High St, Santa Cruz, CA 95064, USA}
\author{Jared R. Males}
\affiliation{Steward Observatory, University of Arizona, 933 North Cherry Avenue, Tucson, AZ 85721, USA}
\author{Katie M. Morzinski}
\affiliation{Steward Observatory, University of Arizona, 933 North Cherry Avenue, Tucson, AZ 85721, USA}

\begin{abstract}
	We present limits on the lens flux of \thisevent\ based on MagAO imaging taken seven years after the microlensing event. At the time of the observations, the lens should have been separated from the microlensing source by 292 mas. However, no new sources are detected with MagAO. We place an upper limit on the lens flux of $H>20.57$. This measurement supports the conclusion of \citet{Gould09} that the lens in this event should be a brown dwarf. This is the first test of a prediction based on the terrestrial microlens parallax effect and the first AO confirmation of a sub-stellar/dark microlens.
\end{abstract}


\section{Introduction}

Measurements of the terrestrial parallax effect in microlensing \citep{HolzWald96,Gould97} should be extremely rare because the radius of the Earth is much smaller than the typical size of the Einstein ring projected onto the observer plane (1--10 au). This effect arises when the separation between two observatories on the Earth is sufficient to produce a measurable difference in the source magnification observed by each observatory. Incidentally, the same conditions that lead to measurable terrestrial parallax effects are optimal for measuring the finite source effect \citep{GouldYee13}. From these two effects one can measure the microlens parallax, $\pi_{\rm E}$, and the Einstein radius, $\theta_{\rm E}$, respectively, and hence, derive the mass of the lens and the distance to the system:
\begin{equation}
    M_{\rm L } = \frac{\theta_{\rm E}}{\kappa \pi_{\rm E}} ; \quad D_{\rm L} = \left((\theta_{\rm E} \pi_{\rm E})\ \mathrm{au} + \frac{\mathrm{au}}{D_{\rm S}}\right)^{-1}
\end{equation}
where $D_{\rm S}$ is the distance to the source (usually the mean distance to the Galactic bulge) and $\kappa \equiv 8.14\ \mathrm{mas}\ \mathrm{M}_{\odot}^{-1}$. \citet{GouldYee13} estimated that only 0.1 mass measurements from the combination of terrestrial parallax and finite source effects should have been made to date (at the time that paper was written). By contrast, two such measurements had been made.

One of these measurements was made for \thisevent\ by \citet{Gould09}\footnote{The other was for OGLE-2008-BLG-279 \citep{Yee09}}. \thisevent\ was magnified by a factor of over 2500, and because of this extreme magnification, both finite source effects and the terrestrial microlens parallax effect were measured. The combination of the finite source effect and the microlens parallax led to the conclusion that the lens was a $0.056 \pm 0.004 M_{\odot}$ thick-disk brown dwarf at a distance of 525 pc with the corresponding prediction that the lens should be very faint: $H=25.7$ mag.

In this paper, we present adaptive optics observations from MagAO that confirm the lens is substellar or non-luminous.

\section{Magellan Adaptive Optics Imaging}

\begin{figure}
    \centering
    \includegraphics[width=0.45\textwidth]{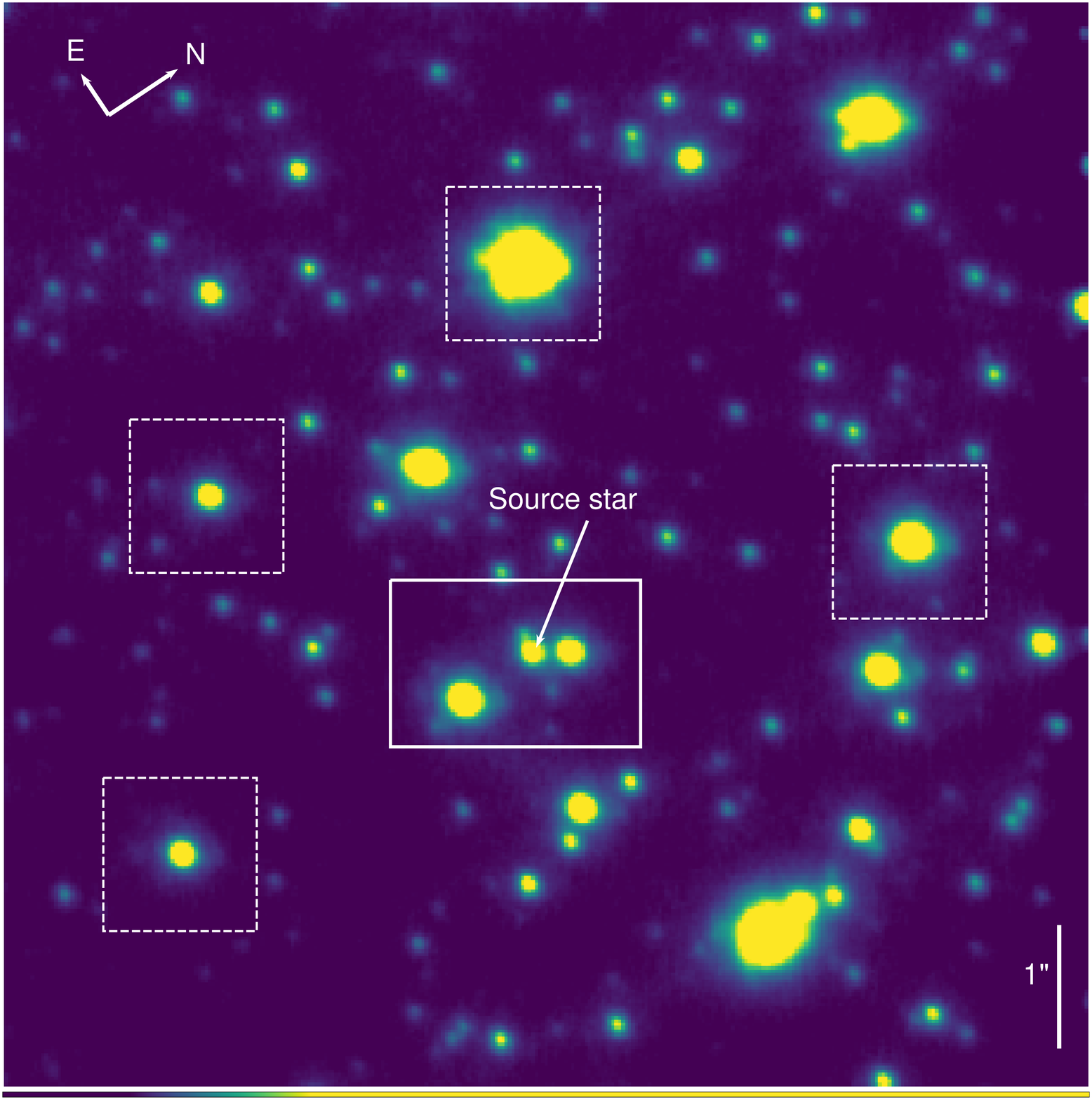}
    \includegraphics[width=0.452\textwidth]{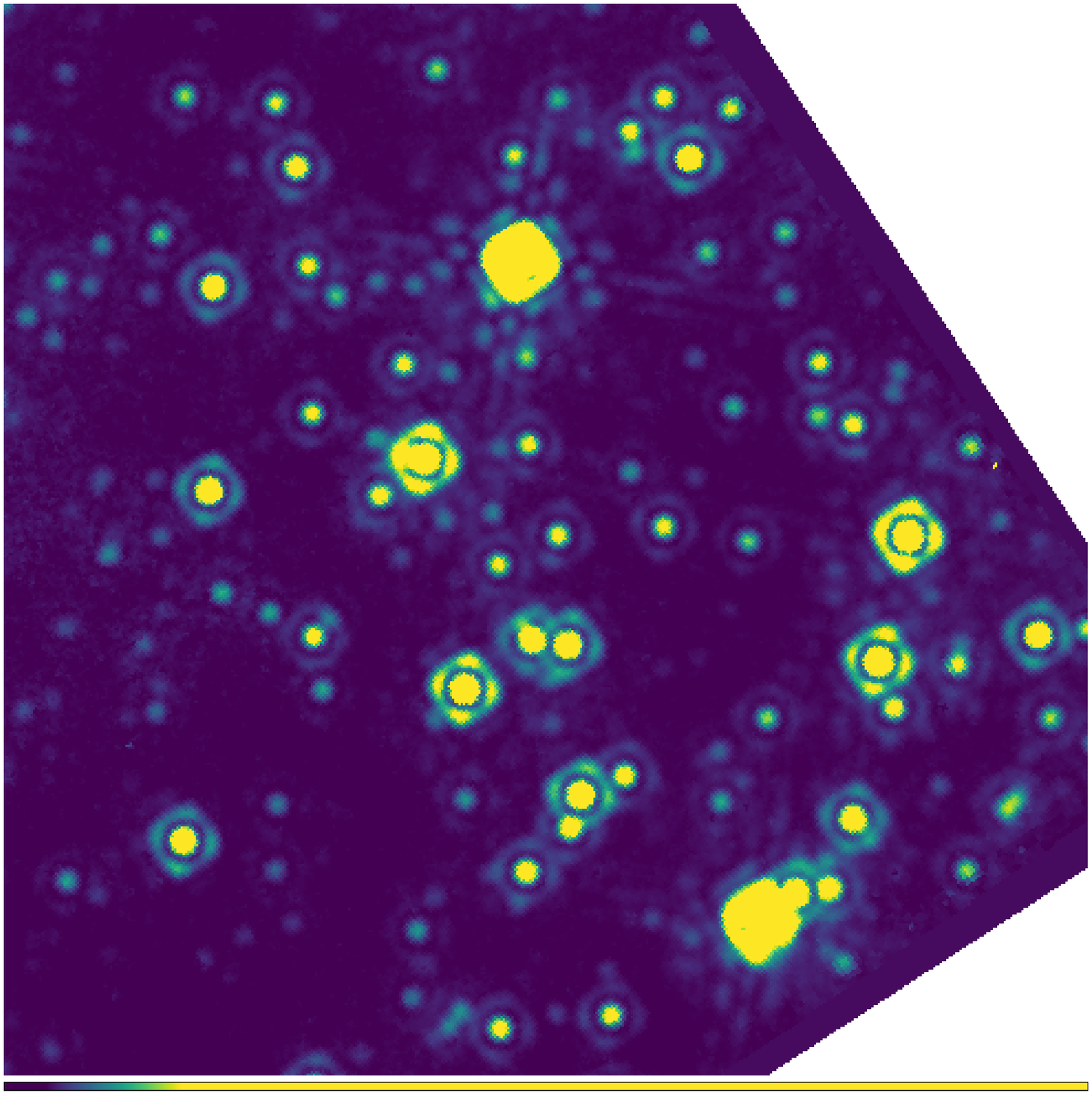}
	\caption{{\em Left}: MagAO imaging from April 2014 of the $8\farcs8 \times 8\farcs8$ field surrounding \thisevent. The position of the microlensing source is indicated by the arrow. The dashed squares mark the stars used for constructing the empirical PSF. The solid rectangle marks the extent of the substamp shown in Figure \ref{fig:magao_subimage}. {\em Right}: HST NICMOS F160W imaging of the same field from June 2008.
    \label{fig:whole_magao_field}}
\end{figure}

\subsection{Observations}

\thisevent\, was observed on the night of 20 April 2014 using the MagAO instrument on the Magellan Clay Telescope \citep{Morzinski16}. For these observations, we used the Clio camera in ``wide" mode, which gives a pixel scale of 27.49 mas pix$^{-1}$ and a $14^{\prime\prime}\times28^{\prime\prime}$ field-of-view. The observations were taken using a four point dither pattern. At each position, five 30s exposures were taken. The observations were made in the $H$ band. A sky field was also observed earlier in the night.

\subsection{Image Processing}

The data were flat-fielded using a difference flat constructed from flats that were taken at twilight on 21 April 2014. They were then sky subtracted using the mean sky frame. Bad pixels were flagged using a custom bad pixel mask, based on pixels with either large standard deviations in the flat frames (``hot" pixels) or low counts (``dead" pixels). After sky subtraction, we found that the median counts for each frame varied as a function of time (by $\lesssim 1\%$ of the raw counts). We measured an offset for each frame and subtracted it. Then, we re-sampled the images at a factor of 5 higher resolution before shifting them and taking the median of the images. The final image is downsampled from this median image to the original resolution.

The location of the target was measured from observations taken with ANDICAM \citep{DePoy03} on the SMARTS 1.3m telescope in Chile, while the event was highly magnified, i.e., on HJD$^{\prime}=$HJD$-2,450,000=4233.6$. These observations were taken in both $I$ and $H$ bands. Only one star in the ANDICAM $H$-band data (other than the target) is bright enough to be detected and also in the MagAO field. 
Thus, we matched the catalog for this frame to the VVV catalog \citep{Minniti17_VVV}, 
and then the VVV catalog to a catalog of stars extracted from the MagAO field using Source Extractor \citep{BertinArnouts96_SEx} to determine the location of the target. We also checked the microlens source position by matching the ANDICAM $I$-band catalog directly to the MagAO field and found consistent results. The identified target matches the description of the field given by \citet{Gould09}.

For the astrometric solution, we used the full $14^{\prime\prime}\times28^{\prime\prime}$ image. However, the two chips of the Clio detector have different photometric responses. Thus, we restricted our flux measurements to the chip containing the science target.

\subsection{Image Analysis}\label{ss:image-analysis}

Because the region surrounding \thisevent\ is relatively crowded (Figures \ref{fig:whole_magao_field} and \ref{fig:magao_subimage}), we fit sources using an empirical PSF constructed from four bright, relatively isolated, stars surrounding the target. The empirical PSF is parameterized by the centroid position (X, Y) and a scaling factor (peak counts). We calibrate the MagAO fluxes by cross-matching to the VVV catalog \citep{Minniti17_VVV}. All of the VVV sources are blends of multiple MagAO stars. For each VVV star, we fit for all sources in a $2\farcs0$ substamp around the target. We then sum the fluxes for all MagAO stars within $0\farcs7$ of the VVV source. In total, there were 8 calibration stars spanning a magnitude range from $H_{\rm VVV} = $ 15.8 to 12.8. From these stars, we derived the flux relation:
\begin{equation}
    H_{\rm VVV} = H_{\rm Clio} + 6.813 \pm 0.015 
\end{equation}
where $H_{\rm Clio}$ is calculated from the peak counts in the PSF (i.e., $H_{\rm Clio} = 18 - 2.5\log_{10}{f_{\rm peak}}$).

To estimate the uncertainties in the measured fluxes, we divided the 20 exposures into two subsets and co-added each subset into new composite images 1 and 2. Then we measured the fluxes of stars in each composite image and took the difference of the two measurement, $\Delta_i = (\mathrm{flux}_{i,1} - \mathrm{flux}_{i,2})$, for each star, $i$. 
In the limit
as the number of stars at a particular apparent magnitude goes to infinity, the uncertainty in the flux in the original co-added image is
\begin{equation}
\sigma_{\rm tot} \equiv \frac{\sqrt{\langle\Delta^2\rangle}}{2}
\label{eqn:sigma_flux}
\end{equation}
for that magnitude. In a real image, there is a finite number of stars spanning a finite set of discrete flux values. So to estimate $\langle\Delta^2\rangle$, we assume that the uncertainty in the flux is a smooth function of the peak flux. We fit a power-law to the measured $\Delta_i^2$ as a function of $\mathrm{flux}_i$ from the original image. Solving Equation \ref{eqn:sigma_flux} for $\sigma_{\rm tot}$ gives us the typical measurement error as a function of flux. 

\section{HST Archival Data}

The field of OGLE-2007-BLG-224 was observed by HST-NICMOS on 11 June 2007 and 8 June 2008 (PI: K. Sahu), respectively about one month and one year after the microlensing event peak. To compare to our MagAO imaging, we retrieved the F160W images taken in 2008 from the HST Legacy Archive. 
We identified the brightest isolated star in the field to use as our $1\farcs\times 1\farcs$ PSF template.

\section{Non-Detection of the Lens}

\begin{figure}
    \centering
	\includegraphics[width=0.5\textwidth]{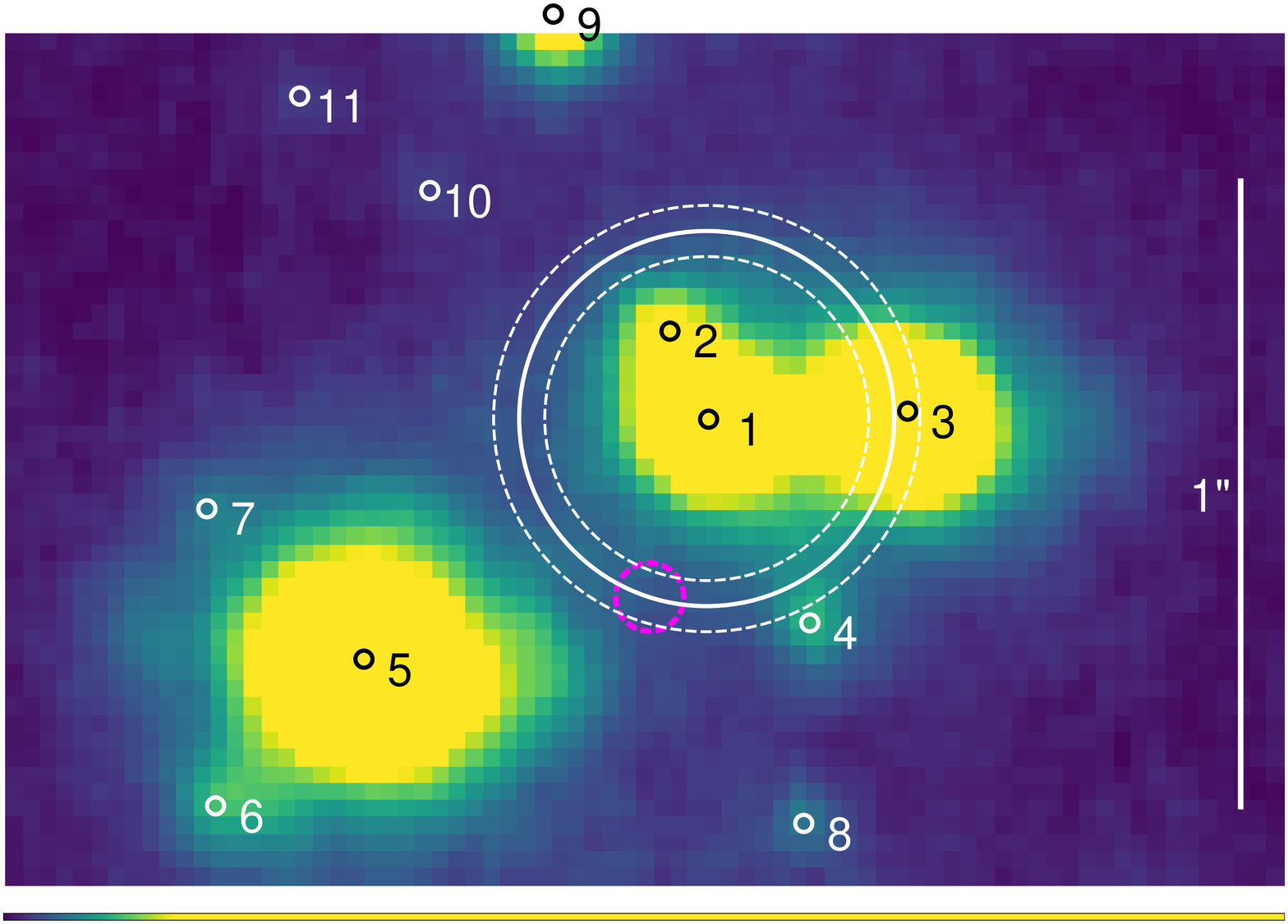}
	\includegraphics[width=0.5\textwidth]{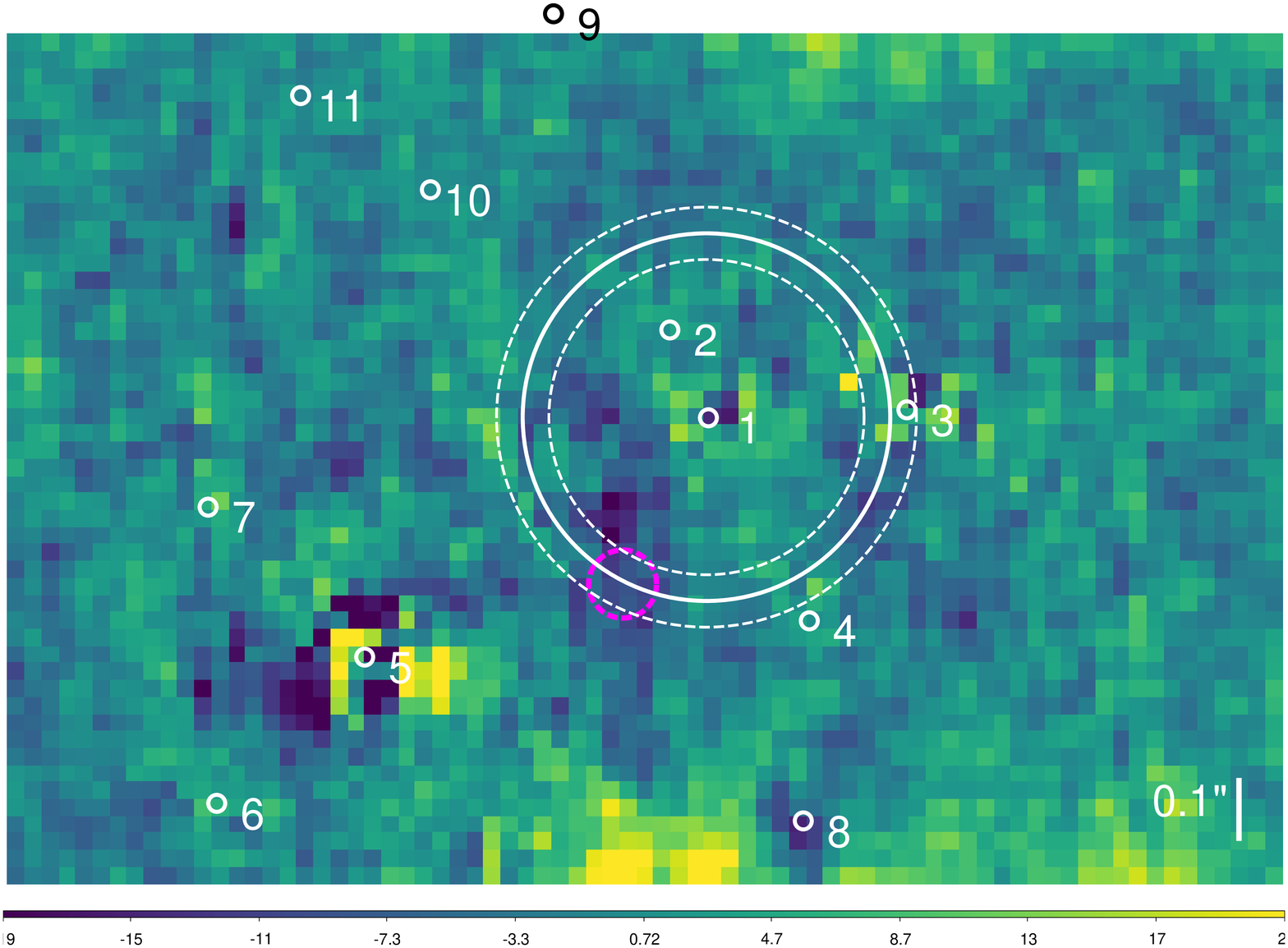}
	\includegraphics[width=0.5\textwidth]{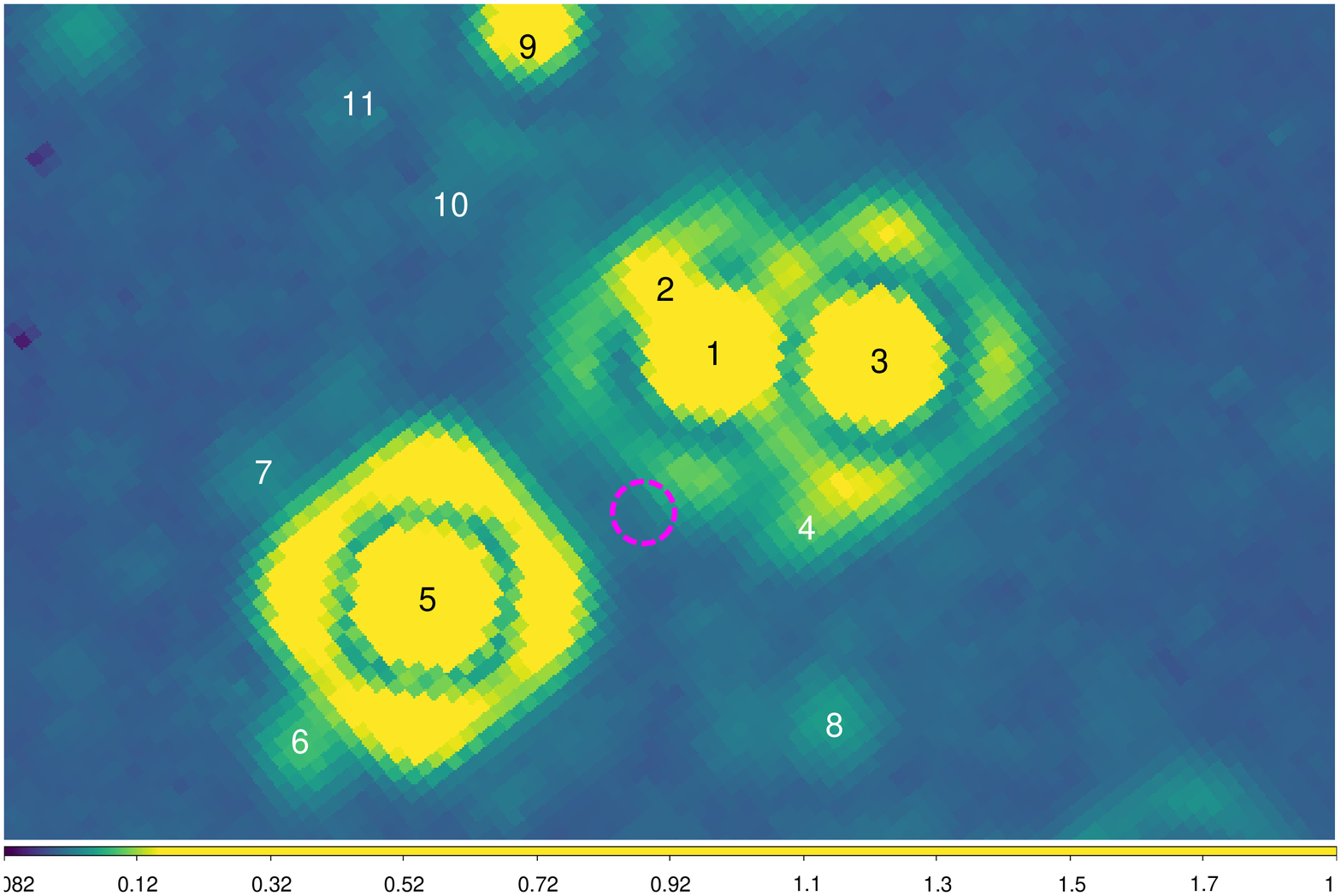}
	\caption{MagAO $2\farcs0 \times 1\farcs3$ substamp showing the microlensing source from \thisevent\ (\#1) and other nearby stars (\#2--11). The large, white circle surrounding the source shows a separation of $292 \pm 14$ mas (solid and dotted lines, respectively), which is the expected separation of the lens from the source based on the magnitude of the proper motion alone. The magenta circle is the expected position of the lens including the microlens parallax measurement of the direction of the proper motion. {\it Top:} MagAO from 2014, {\it middle:} residuals to the MagAO image after PSF fitting (the colorbar ranges from $-20$ to $20$ counts), {\it bottom:} HST F160W imaging from 2008. The HST PSF has a ring feature corresponding to the first Airy ring. \label{fig:magao_subimage}}
\end{figure}

Based on the analysis of \citet{Gould09}, we expect the lens to be a brown dwarf and, therefore, faint and undetectable in the MagAO images. We can predict the location of the lens based on the lens-source relative proper motion measured by \citet{Gould09}:
\begin{eqnarray}
    \mu_{\rm hel} & = & \tilde{v}_{\rm hel} \left(\frac{\pi_{\rm rel} }{\mathrm{au}}\right) = 42.06 \pm 2\ \mathrm{mas\ yr}^{-1}\\
    \hat{\mu}_{\rm hel} & = & 61^{\circ} \pm 5^{\circ}\ \mathrm{S\ of\ W}.
\end{eqnarray}
Thus, at the time of the MagAO observations the lens and microlensing source are expected to have separated by $292 \pm 14$ mas. 

The expected location of the lens is shown in Figure \ref{fig:magao_subimage}. The magenta circle shows the predicted location of the lens, including the measurement of the direction of the proper motion from the terrestrial parallax. The white annulus shows the constraints on the predicted lens location based on the measurement of $\theta_{\rm E}$ alone. If the terrestrial parallax measurement were in error, and the lens were a main sequence star, we would expect an additional source to be detected somewhere along this annulus.

\subsection{Fitting for Known Sources}

\begin{figure}
	\includegraphics[width=\textwidth]{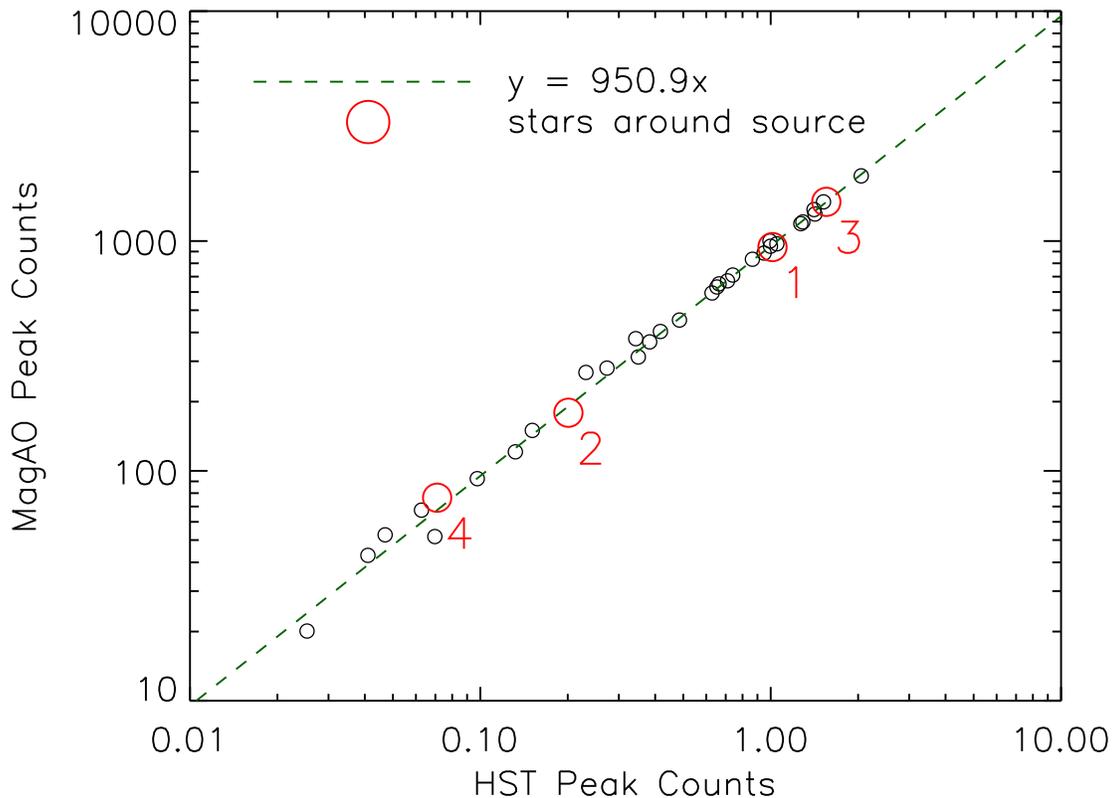}
	\caption{Fluxes for stars measured in the MagAO image are well-matched to fluxes measured from the 2008 HST image. In particular, no significant deviation is seen for the stars closest to the microlensing source (red circles, numbered as in Figure \ref{fig:magao_subimage}). \label{fig:hst_fluxes}}
\end{figure}

\begin{table*}[htb]
\begin{center}
\caption{Properties of Sources Near \thisevent}
\label{tab:subimage_sources}
\begin{tabular}{ccccccc}
\hline\hline
Source ID & $X_{\rm centroid}$ & $Y_{\rm centroid}$ & $H_{\rm 2MASS}$ \\
 & (pixels) & (pixels)  &  & & &\\
\hline
         1  & 346.790  & 298.883  &   17.38$\pm$0.04 \\
         2  & 344.522  & 304.047  &   19.18$\pm$0.07 \\
         3  & 358.467  & 299.352  &   16.89$\pm$0.03 \\
         4  & 352.726  & 286.942  &   20.10$\pm$0.11 \\
         5  & 326.571  & 284.812  &   16.10$\pm$0.03 \\
         6  & 317.868  & 276.211  &   20.28$\pm$0.12 \\
         7  & 317.360  & 293.629  &   21.29$\pm$0.20 \\
         8  & 352.372  & 275.179  &   20.07$\pm$0.11 \\
         9  & 337.679  & 322.640  &   18.59$\pm$0.06 \\
        10  & 330.436  & 312.293  &   21.62$\pm$0.24 \\
        11  & 322.803  & 317.831  &   21.37$\pm$0.21 \\
\hline\hline
\end{tabular}
\end{center}
Note: Sources are numbered as in Figure \ref{fig:magao_subimage}\\
\end{table*}

We begin by fitting for sources in a rectangular ($2\farcs0 \times 1\farcs3$) substamp (Figure \ref{fig:magao_subimage}) containing the source star of the event and its immediate neighbors. These fits use the empirical PSF described in Section \ref{ss:image-analysis} plus a constant background level for the image. Sources were recovered iteratively by examining the residuals to the fitted image and comparing them to the HST NICMOS images of the field. In total, 11 sources were recovered in the ($2\farcs0 \times 1\farcs3$) substamp. All of them can be cross-matched to stars in the HST NICMOS imaging. Their properties are given in Table \ref{tab:subimage_sources}. The residuals to the eleven-source fit are shown in Figure \ref{fig:magao_subimage}. There is no indication of an additional source within the white annulus.

The one exception to this could be if the lens were very near to source \#3, which happens to lie at the expected lens-source separation. In that case, light from the lens might be absorbed into the fits for source \#3. To check for this possibility we compare the fluxes of the stars to the fluxes measured from the HST image from 2008 (Figure \ref{fig:hst_fluxes}). The fluxes for all the sources are consistent between MagAO and HST, with no significant excesses. Therefore, we conclude that source \#3 is not hiding significant light from the lens.

Finally, from the MagAO image, we measure $H = 17.38 \pm 0.04$ for the source at the position of \thisevent\ (source \#1). This is slightly brighter than the microlens source flux as reported in \citet{Gould09} ($H_S = 17.50 \pm 0.02$), which is marginally inconsistent at $2.7 \sigma$. The excess would correspond to an $H=19.8$ star. However, the measured source flux is also consistent with the flux measured for the same source in the HST imaging.
\citet{Gould09} state that the HST source is consistent with zero additional flux aside from the microlens source. Even if the excess is real, it is 
in the wrong place to be the lens, although it could be a companion to the microlens source. 

\subsection{Search for Additional Sources}

To confirm that there are no additional sources in the field, we conduct a blind search. We seed fits for a twelfth source at every pixel over a $1\farcs35 \times 1\farcs35$ square surrounding \thisevent S. The peak flux for the PSF is initialized at 50 counts, and then this source is fit freely simultaneously with the eleven known sources. We consider a source detected if the final fit is inside the search grid, within 20 pixels of the starting location, and has a fitted peak flux of at least 10 counts. We tested our method for eleven-star fits to confirm that the two faintest sources (10 and 11) are recovered.

There are two classes of ``detections" that survive these cuts. First, there are ``detections" very close to the centroids of known sources in the field. In these cases, the twelfth source partially supplants the known source; the sum of their fluxes remains the same as for the known source alone (in particular, this is true for source \#3, i.e., there is no evidence for an additional source ``hidden" in the wings of this star). Hence, we apply an additional cut that the new detection should be at least 1 FWHM from a previously known source to eliminate this class of ``detections." Second, there are ``detections" of a twelfth source along the bottom edge of the substamp. 
There are unexplained residuals in this region at the level of $\sim 20$ counts, possibly from unresolved faint sources, associated with the wings of bright sources outside the substamp, or simply spurious. However, because this string of ``detections" are also separated from the source by almost twice the expected distance travelled by the lens, even if they are real, they are unlikely to be the lens. 

Thus, this blind search did not reveal any new sources that are plausible candidates for the lens of \thisevent.
 
\subsection{Limits on Additional Stars}

To assess the limits on non-detections in the image, we inject fake sources into the image and attempt to recover them. Fake sources are injected at every pixel over the $1\farcs35\times1\farcs35$ grid used for the blind search. Then, fits are seeded on a $3\times3$ `mini-grid' (with spacings between points of 2.5 pixels) around the injected source. A source is considered to be recovered if more than 5 out of the 9 fits initialized from the mini-grid around the injected source meet the following conditions: 1) the retrieved centroid is less than 1 pixel from the injected location and 2) the retrieved peak flux is within 50\% of the injected value. Thus, we can place a detection sensitivity limit at every grid point around the microlensing source star. Note that this detection criterion is considerably more stringent than that used for the blind search. Hence, it results in a conservative lower limit on the brightness of recoverable sources at each location.

At the expected location of the lens, we find the $3\sigma$ flux limit to be $H > 20.57$. Along the annulus defined by $\mu_{\rm rel, hel}$, the limits are generally better except where the annulus intersects the PSF of stars 3 and 4. Although the limits nominally reach $H \sim 22.1$ in some places, these limits are highly dependent on the exact level of the background in a particular pixel, and thus, may be over-optimistic. However, stars 10 and 11 (with magnitudes $H = 21.6$ and $H = 21.4$, respectively) are clearly recovered in the blind search. Hence, we can conservatively say that the limits on the lens flux range from 21.6 to 20.6 depending on the location along the annulus.  

\section{Discussion}

\begin{figure}
    \centering
	\includegraphics[width=\textwidth]{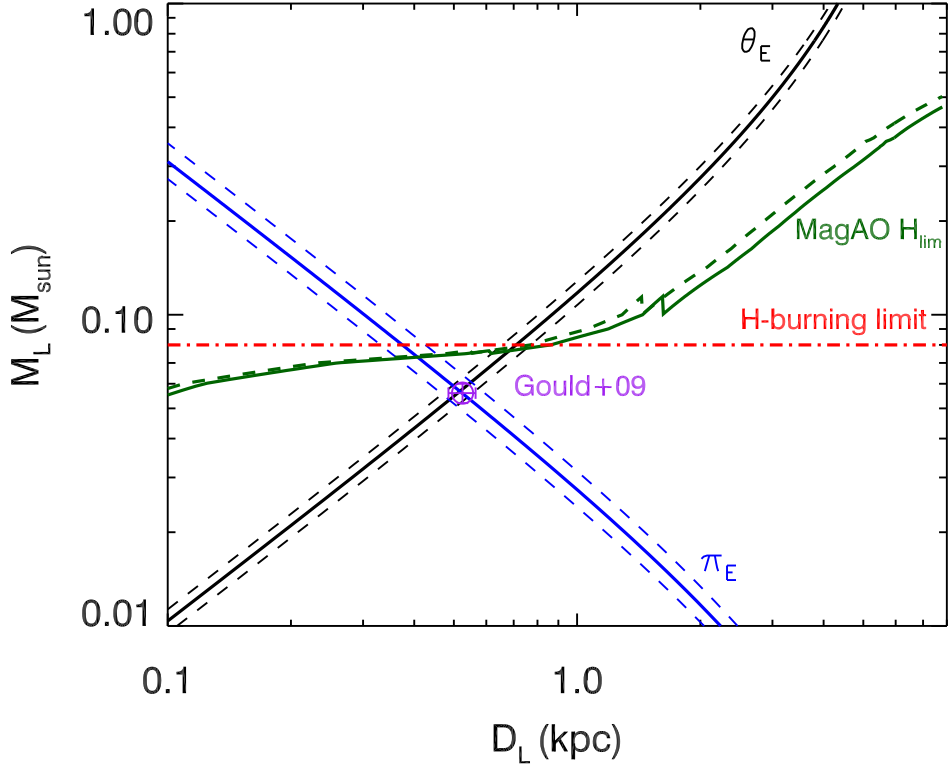}
	\caption{Constraints and limits on the mass and distance of \thisevent L. The constraints (solid lines) from $\theta_{\rm E}$ and $\pi_{\rm E}$ from \citet{Gould09} are shown in black and blue, respectively, with their $1\sigma$ limits (dashed lines). The green line shows the flux limits from MagAO after transforming to mass and distance using stellar and sub-stellar isochrones. The dashed green line shows the limit if an extinction of $A_H = 0.25$ is included. Lenses to the upper left of the green line are excluded. This limit clearly shows that only substellar lenses (dot-dashed red line) are consistent with the microlensing measurements.
    \label{fig:mldl}}
\end{figure}
	
We have found no evidence for a luminous lens in \thisevent, and we have placed upper limits on light from the lens of $H > 20.57$ mag. These limits are consistent with the prediction from \citet{Gould09} that the lens should be $H =25.7$. We transform this flux constraint into mass and distance using 5-Gyr, solar-metallicity isochrones from MIST \citep{Choi16} for $> 0.1 M_\odot$, BHAC15 \citep{Baraffe15} for $0.075$ to $0.1 M_\odot$, and \citet{Baraffe03} for $< 0.075 M_\odot$ (note that because the isochrones are not constrained to be consistent with each other, this results in a slight hitch in the mass-luminosity relation.). In this field, \citet{Gonzalez11,Gonzalez12} measure $E(J-K) = 0.2561$ and $A_K = 0.1353$, so we adopt $A_H = 0.25$ as a reasonable estimate for the  $H$-band extinction toward this field. 
However, the value of the extinction and whether or not the lens experiences some or all of it has very little effect on our conclusions. Figure \ref{fig:mldl} shows that our limit strongly constrains the lens mass and distance, independent of any other information. It also compares this limit to the constraints on the lens mass and distance from $\theta_{\rm E}$ and $\pi_{\rm E}$ assuming source distance of 7.8 kpc \citep{Gould09}. The lack of a detection of light from the lens in \thisevent\ strongly supports the measurement of the lens mass and distance based on the finite source and terrestrial parallax effects measured from the light curve. This is the first test of a terrestrial parallax measurement.

Recently, there have been several discoveries of strong free-floating planet candidates \citep{Mroz17FFPs,Mroz18,Kim20} including a Mars- to Earth-mass object \citep{Mroz20FFP}. Proving that these planets are truly free-floating requires demonstrating that they do not have host stars. Stellar hosts at several to $\sim 10\ $ au for these candidates have been ruled out by the microlensing light curves but could exist at wider separations. The limits on lens light for \thisevent\ demonstrate that current instruments can place flux limits strong enough to rule out stellar companions to these candidates.

\acknowledgements{
We thank A. Gould for comments on the manuscript and M. van den Berg for identifying sky fields for the MagAO observing run. 
KM's work was supported by the NASA Exoplanets Research Program (XRP) by cooperative agreement NNX16AD44G.
This work makes use of observations made with the NASA/ESA Hubble Space Telescope, and obtained from the Hubble Legacy Archive, which is a collaboration between the Space Telescope Science Institute (STScI/NASA), the Space Telescope European Coordinating Facility (ST-ECF/ESA) and the Canadian Astronomy Data Centre (CADC/NRC/CSA). It also
makes use of data products from observations made with ESO
Telescopes at the La Silla Paranal Observatory under programme ID
179.B-2002.
Computations in this paper were run on the FASRC Cannon cluster supported by the FAS Division of Science Research Computing Group at Harvard University.}

\end{document}